\PassOptionsToPackage{pdfpagelabels=false}{hyperref}
\documentclass[fleqn,usenatbib]{mnras}

\usepackage{times} 
\usepackage{epsfig}
\usepackage{amsmath}
\usepackage{graphicx}
\usepackage{footmisc}
\usepackage[usenames,dvipsnames]{xcolor}
\usepackage{natbib}

\newcommand{\dd}{{\rm d}}

\newcommand{\fett}[1]{\boldsymbol{#1}}
\newcommand{\be}{\begin{equation}}
\newcommand{\ee}{\end{equation}}
\newcommand{\deltaGR}{\delta_{\text{\tiny GR}}}

\usepackage[dvipsnames]{xcolor}
\definecolor{darkgreen}{rgb}{0,0.5,0}
\definecolor{babypink}{rgb}{0.78,0.08,0.52}

\makeatletter
\def\@fnsymbol#1{\ensuremath{\ifcase#1\or \mbox{${^{\star}}$}\or *\or \dagger\or \ddagger\or
   \mathsection\or \mathparagraph\or \|\or **\or \dagger\dagger
   \or \ddagger\ddagger \else\@ctrerr\fi}}
\makeatother

\title[COSIRA: Hybrid simulations with radiation]{Cosmological $N$-body simulations including radiation perturbations}

\author[Jacob Brandbyge {\it et al.}]
{%
 Jacob Brandbyge,$^{1,2}$
 Cornelius Rampf,$^{3,4}$\thanks{E-mail: rampf@physics.technion.ac.il}  
 Thomas Tram,$^5$ 
 Florent Leclercq,$^5$ \newauthor 
 Christian Fidler$^6$ and Steen Hannestad$^1$
  \\
 $^1$ Department of Physics and Astronomy, University of Aarhus, Ny Munkegade 120, DK--8000 Aarhus C, Denmark \\
 $^2$ Centre for Star and Planet Formation, Niels Bohr Institute \& Natural History Museum of Denmark, University of Copenhagen, \\~~~~{\O}ster Voldgade 5-7, DK--1350 Copenhagen, Denmark\\
 $^3$ Department of Physics, Israel Institute of Technology -- Technion, Haifa 32000, Israel \\
 $^4$ Institute for Theoretical Physics (ITP), Philosophenweg 16, Heidelberg University,  D--69120 Heidelberg, Germany \\
 $^5$ Institute of Cosmology and Gravitation, University of Portsmouth, Portsmouth PO1 3FX, United Kingdom \\
 $^6$ Catholic University of Louvain -- Center for Cosmology, Particle Physics and Phenomenology (CP3), \\~~~~Chemin du Cyclotron 2, B--1348 Louvain-la-Neuve, Belgium
}

\begin{document}                 
 
\date{\today}

\pagerange{1-5} \pubyear{2016}

\maketitle

\begin{abstract}
Cosmological $N$-body simulations are the standard tool to study the emergence of the 
observed large-scale structure of the Universe.  
Such simulations usually solve for the gravitational dynamics of matter within the 
Newtonian approximation, thus discarding general relativistic effects such as the 
coupling between matter and radiation ($\equiv$ photons and neutrinos). 
In this paper we investigate novel hybrid simulations 
which incorporate interactions between radiation and matter to the leading order 
in General Relativity, whilst
evolving the matter dynamics in full non-linearity according to Newtonian theory.
Our hybrid simulations come with a relativistic space-time and 
make it possible to investigate structure formation  in a unified framework.
In the present work we focus on simulations initialized at $z=99$, and show that
the extracted matter power spectrum receives up to $3\%$ corrections on very large scales through radiation.
Our numerical findings compare favourably with linear analytical results from
\citet{Fidler:2016tir}, from which we deduce that there cannot be any
significant non-linear mode-coupling induced through linear radiation corrections.
\end{abstract}

\begin{keywords}
cosmology: theory -- large-scale structure of Universe -- dark matter
\end{keywords}



\section{Introduction}

According to the $\Lambda$CDM model, we live in a Universe that is nowadays dominated by a cosmological constant ($\Lambda$) and cold dark matter (CDM), whereas the impact of baryons and radiation (i.e., photons and neutrinos) is secondary. None the less, 
at sufficiently early times, baryons and radiation played yet a major role in the early
gravito-electroweak dynamics, 
and were for example responsible for the observed acoustic oscillations in the cosmic microwave background (CMB; \citealt{Smoot:1992td,Netterfield:2001yq}).
Cosmological structure formation is mainly the result of gravitational instability, with initial conditions set in the period of recombination which is around 380,000 years after the Big Bang.
Electroweak interactions freeze out at recombination, and baryons are released from their tight coupling to radiation. Subsequently, this freeze-out in interactions enables matter to cluster significantly for the first time.

At early times such as recombination, a Newtonian approximation of structure formation is  
not appropriate. Instead, the evolution of the  multi-fluid components
(dark matter, baryons, neutrinos, and photons)
is governed by the coupled set of Einstein--Boltzmann equations, i.e., General Relativity (GR).
At early times, the {\it linearized} Einstein--Boltzmann equations
are an excellent description, but
cosmological structure formation for matter becomes fairly quickly 
a non-linear problem, and, unfortunately, solving the 
coupled set of Einstein--Boltzmann 
equations in full non-linearity is not yet feasible. 
Instead, it is common to solve for the process of cosmological structure formation 
using cosmological $N$-body simulations \citep{Stadel2001thesis,Teyssier:2001cp,Springel:2005mi}.  
Such simulations usually demand the Newtonian approximation and evolve only the matter component. 
Therefore, these simulations neglect the evolution of radiation perturbations.

Recently, the first general relativistic cosmological simulations have been performed, either by using an $N$-body approach in the weak-field approximation \citep{Adamek:2014xba,Adamek:2015eda}, or by assuming the fluid approximation for a pure CDM component in full  numerical relativity \citep{Bentivegna:2015flc,Giblin:2015vwq,Mertens:2015ttp}. While  the former includes also (massive) neutrinos, generally these simulations do not take into account the evolution of all multi-fluid species, and furthermore also rely on the validity regime of the weak-field and fluid approximation, respectively. Here we follow a different strategy which comes, of course, with different approximations and assumptions. We present a modification to the Newtonian $N$-body code \texttt{GADGET-2} such that it solves for the full multi-species to first order in cosmological perturbation theory, whilst evolving the matter component in full non-linearity. Essentially, to achieve this modification we use a 
first-order Einstein--Boltzmann solver to compute the force exerted on matter through radiation, and update the momentum conservation of matter in the simulation
accordingly. By pairing the Einstein--Boltzmann solver with the $N$-body code, we 
obtain hybrid simulations with mutual benefits whilst minimizing the
disadvantages from both Newtonian and relativistic worlds. First and foremost, cosmological $N$-body codes in the Newtonian approximation have been tested and improved over a long time period (e.g., \citealt{Schneider:2015yka}), which gives us confidence that the non-linear matter is evolved to 
sufficient accuracy. Secondly, the $N$-body output of our relativistic 
simulations is in accordance with first-order cosmological perturbation theory, 
and thus comes with a solid approximation of the underlying relativistic space-time.
This should serve as a solid basis for e.g.\ investigating ray tracing.

This paper is organized as follows. In the following section we report the relativistic fluid equations, which we aim to solve numerically by modifying the exisiting $N$-body code \texttt{GADGET-2}. We pair the simulation with the Einstein--Boltzmann solver \texttt{CLASS}, and call the resulting code \texttt{COSIRA}.
 Details on the specific implementation can be found in
Section \ref{sec:numerics}. 
We discuss our numerical results and confront them with theoretical predictions in  Section~\ref{sec:results}. 
Finally, we summarize and discuss our results in Section~\ref{sec:conclusion}.

\section{Equations of motion in General Relativity}\label{sec:EoMs}

In a {\it multi-fluid} Universe,  the relativistic equations of motion for CDM 
in an arbitrary gauge contain general relativistic corrections at multiple instances. 
Here we work in the $N$-body gauge which \textit{minimizes} the appearance of 
such GR corrections.
As was shown in \citet{Fidler:2015npa}, a fully relativistic analysis of the 
multi-fluid dynamics reveals, 
that momentum and mass conservation for CDM are to 
the leading order given by
\begin{subequations}
\begin{align}
  &\partial_\eta \fett{v}_{\rm cdm} + {\cal H}\, \fett{v}_{\rm cdm} \hspace{0.015cm}= \nabla\Phi +  \nabla\gamma \,, \label{eq:Euler} \\
 &\partial_\eta \delta_{\rm cdm} + \nabla\! \cdot \!\fett{v}_{\rm cdm} =  0 \,, \label{eq:conti}
\end{align}
respectively.
Here, $\eta$ is the conformal time, ${\cal H} = \dot a/a$ the conformal Hubble parameter with $a$ being the cosmic scale factor, 
$\fett{v}_{\rm cdm}$ and $\delta_{\rm cdm}$ are respectively the (peculiar) velocity and the density contrast of the dark matter component; $\Phi$ is the cosmological potential satisfying the Poisson equation
\begin{align}
  &\nabla^2 \Phi =  - 4\pi G  a^2 \sum_\alpha \bar \rho_\alpha \delta_\alpha \,, \label{eq:Poisson}
\end{align}
\end{subequations}
where $\bar \rho_\alpha$ is the background density of the species labelled with $\alpha$, and the summation on the r.h.s.\ runs over all relevant species in the Universe, i.e., dark matter, baryons, neutrinos and photons.

The above equations have been derived from GR in the $N$-body gauge with corresponding line element
\be \label{lineNbody}
   \dd s^2 \!= \! a^2 \Big[ - (1+2 A ) \dd \eta^2  -2  v_i \dd x^i \dd \eta  \              + \left( \delta_{ij} - 2 D_{ij} H_{\rm T} \right) \dd x^i \dd x^j \Big] ,
\ee
where $D_{ij} \equiv \partial_i \partial_j - \delta_{ij} \nabla^2/3$, $A$ is a perturbation in the time-time component of the metric  ($A$ is sourced by radiation pressure and anisotropic stress), $v_i = \partial_i v$ is the scalar part of the total velocity of all fluid components, where $v$ is the potential of the said scalar part, and $H_{\rm T}$ is a perturbation in the trace-free part of the spatial metric. These linear perturbations are determined by GR and 
can be obtained from conventional Einstein--Boltzmann codes. 
Also, the relativistic correction
\begin{align} \label{gamma}
  \gamma \equiv   \ddot H_{\rm T} + \frac{\dot{a}}{a}\dot{H}_{\rm T} 
  - 8 \pi G a^2 \sum_{\alpha}  p_\alpha \Pi_\alpha  
\end{align}
can be obtained from such codes. Here, $p_\alpha$ and $\Pi_\alpha$ are respectively the pressure and anisotropic stress of photons and neutrinos.

\begin{figure}
\includegraphics[width=\columnwidth]{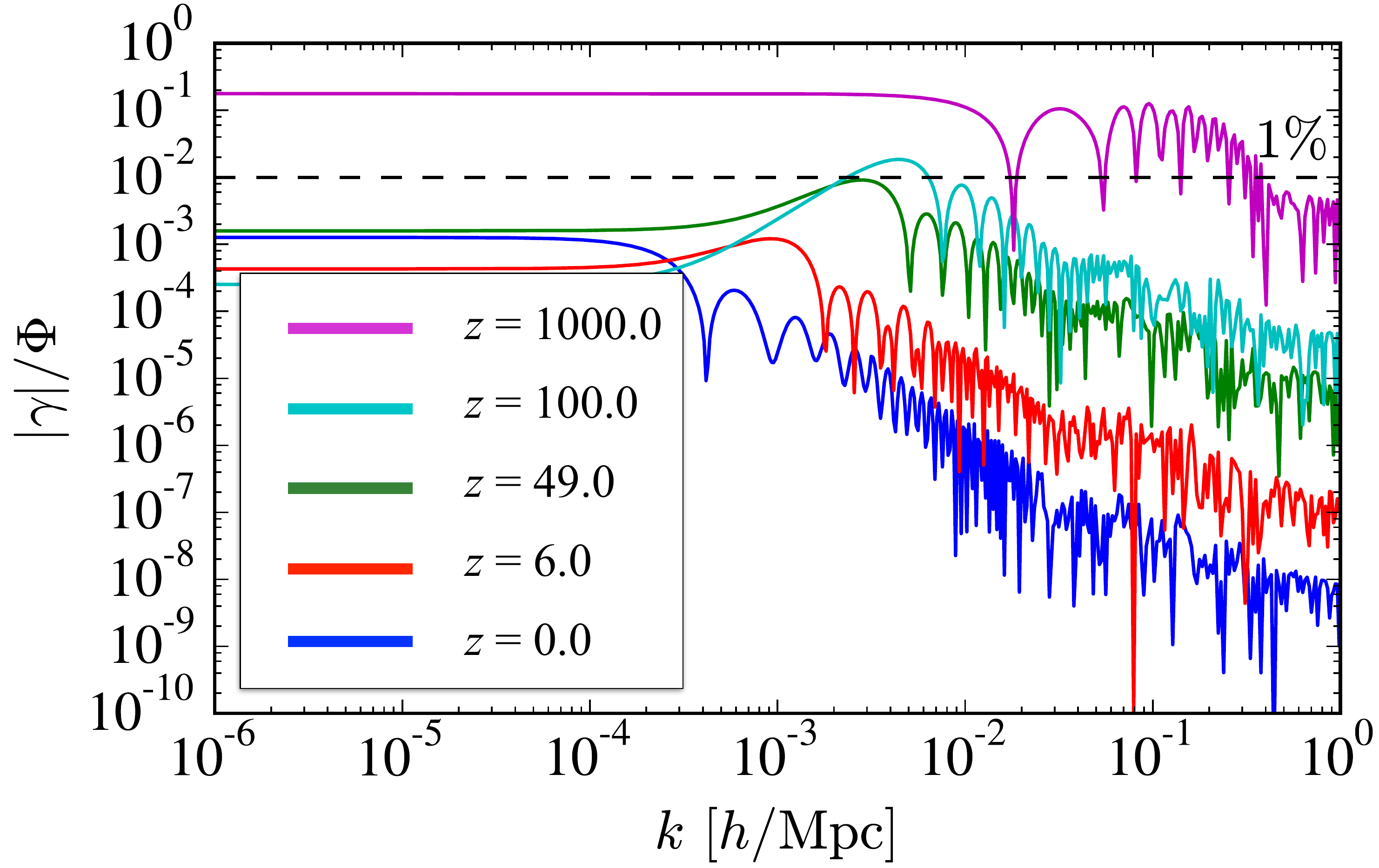}
\caption{\label{fig:Phiratio} 
Ratio of $|\gamma|$ compared to the cosmological potential $\Phi$ in Fourier space, illustrating the impact of radiation contaminants on the momentum conservation equation~(\ref{eq:Euler}) of the dark matter component. 
On scales $k \geq 10^{-3}h/$Mpc the relevance of residual radiation is continuously decreasing in time. Figure taken from \citet{Fidler:2015npa}.
}
\end{figure}

Equations~(\ref{eq:Euler})--(\ref{eq:Poisson}) are, {\it apart from the additional function $\gamma$ on the r.h.s.\ in}~(\ref{eq:Euler}) \textit{and the non-CDM source terms
in}~(\ref{eq:Poisson}), identical with the Newtonian fluid equations.
The function $\gamma$ is non-zero when there is a significant amount of radiation, whereas it is vanishing when radiation becomes negligible. 
Chronologically, this is precisely the case during cosmological structure formation.
As can be seen from Fig.\,\ref{fig:Phiratio}, at early times ($z \geq 50$) the $\gamma$ term yields corrections of up to $1\%$ -- $20\,\%$ to the momentum conservation
 of the CDM component, whereas at late times ($z<50$), $\gamma$ becomes 
negligible on all scales.
Furthermore, at earlier times when there is still significant radiation, the Poisson equation~(\ref{eq:Poisson}) is sourced by non-vanishing density perturbations from 
all species, whereas at late times it is effectively sourced only by matter perturbations.
Summing up, equations~(\ref{eq:Euler})--(\ref{eq:Poisson}) coincide with the Newtonian fluid equations to the leading order, but only at late times when $\gamma$ and the non-matter density perturbations are vanishing.
Thus, only in that limit, conventional Newtonian $N$-body simulations solve for the dynamics of dark matter in accordance with GR.  
In the following section we will introduce a modification to $N$-body simulations which makes it possible to run conventional simulations at early times, and in accordance \mbox{with GR.}
Having implemented the relativistic corrections in the Newtonian simulation, the respective $N$-body output satisfies automatically first-order GR for all fluid species,
and thus comes with an approximation of the underlying space-time, here to be interpreted in the $N$-body gauge.

\section{numerical implementation}\label{sec:numerics}

While conventional $N$-body simulations solve for the gravitational collapse of CDM
in full non-linearity, they do not incorporate relativistic terms which stem from the 
presence of other fluid species. Here we introduce the code \texttt{COSIRA}, standing 
for COsmological SImulations with RAdiation. \texttt{COSIRA} makes it possible to perform hybrid simulations which are compatible with General Relativity.

To achieve this, we separate the force term in the momentum conservation for CDM 
(see r.h.s.\ of equation~(\ref{eq:Euler}))
 into parts which are determined either in the $N$-body simulation or in the Einstein--Boltzmann code, 
\begin{subequations}
\be
  \Phi + \gamma = \Phi_{\rm sim} + \Phi_{\text{\tiny GR}} \,,
\ee
where 
\begin{align}
  &\nabla^2 \Phi_{\rm sim}\!=  - 4\pi G  a^2  \bar \rho_{\rm cdm} \delta_{\rm sim} \,, \\
  &\nabla^2 \hspace{0.02cm}\Phi_{\hspace{0.014cm}\text{\tiny GR}}\,  = - 4\pi G  a^2  \bar \rho_{\rm cdm} \deltaGR \,, \label{def:deltaGR}
\end{align}
\end{subequations}
where \mbox{$\bar \rho_{\rm cdm} \deltaGR \equiv \sum_{\alpha\neq \rm cdm} \bar \rho_\alpha \delta_\alpha - (4\pi Ga^2)^{-1}\nabla^2 \gamma$}, 
and $\delta_{\rm sim}$ is the non-linear CDM density contrast obtained from 
the $N$-body simulation.
The \textit{effective} GR perturbations, $\deltaGR$,
are determined by using a modified version of the Einstein--Boltzmann 
code \texttt{CLASS} \citep{Blas:2011rf},  
and this information is fed to the Poisson solver in the $N$-body simulation at each time step.
We note that in the present approach, for reasons of consistency the information is only passed from the Einstein--Boltzmann system to the $N$-body code, but not vice versa. This means that we include the linear impact of the relativistic species 
on the non-linear evolution of CDM, but not the backreaction of the non-linear clustering of matter on the metric potentials or the relativistic species.

As evident from Fig.\,\ref{fig:Phiratio}, $\gamma$ is rapidly
oscillating in $k$ (over most scales) and therefore also oscillates in time. 
Since the time-scale of these oscillations is very short, of order $1/k$, it is not feasible to track them in an $N$-body simulation. However, as was shown in
 \citet{Weinberg:2002kg} and more recently in \citet{Voruz:2013vqa},  CDM dos not couple to these fast modes, and thus we consider only the slowly varying mode of $\gamma$ in our implementation. 
To extract this averaged mode of $\gamma$, we first apply a cutoff of $k \leq 3 \cdot 10^{-3}\,h/$Mpc  below which an average procedure is not required. For $k$-scales larger than the cutoff,
we resample the transfer function of $\gamma$  such that it has 40 points per period $T=2\pi/k$. Then we compute the rolling mean with a window size of 40 points (i.e., exactly one period). This rather complicated method is needed as at early times the oscillator is driven, implying that the frequency is not constant. 

Our hybrid simulations have a box size of $L = 16384$\,Mpc$/h$ and a resolution of 1024 particles per dimension. The initial conditions for the CDM component are given in terms of the Zel'dovich approximation \citep{Zeldovich,Buchert:1992ya}, and are initialized with a weighted CDM and baryon power spectrum at $z_{\rm ini} =99$, which can be obtained from e.g.\ \texttt{CAMB} \citep{Lewis:1999bs}.
For the cosmological parameters, we use  $\Omega_{\rm m} = 0.3133$, $\Omega_{\Lambda} = 0.6867$, $\Omega_{\rm b} = 0.0490$, $h = 0.6731$, $n_{\rm s} = 0.9655$, $A_s=2.215\cdot 10^{-9}$,  $\sigma_8 = 0.845$, and we consider a cosmology with three massless neutrino species.

In the present work, the effective GR perturbations are not represented by discretized $N$-body particles but in terms of a fluid description.  These GR perturbations are updated whenever a long-range force is calculated in \texttt{GADGET-2}, and then added to the CDM long-range perturbations in Fourier space.
Since there is no small-scale tree part for the radiation perturbations, they are, unlike the CDM perturbations, not smoothed with a Gaussian factor.
For more details we refer to \citet{Brandbyge:2008js} where similar techniques have been applied to include linear perturbations in $N$-body simulations, however in a Newtonian set-up with massive neutrinos. 

Note also that for reasons of better comparison with theoretical predictions, we use the same initial matter power spectrum and set of phases, 
for both runs with or without GR perturbations. Furthermore,
to account for the radiation in the background evolution of the simulations, we added the background component of the radiation density to the Friedmann equation in \texttt{GADGET-2} in both simulations with and without GR perturbations.

CDM density fields are obtained by assigning $N$-body particles to grid points by applying the cloud-in-cell (CIC) algorithm \citep{CIC}. When plotting pure estimates of the CDM density (right panel in Fig.\,\ref{fig:relativedifferencedelta}), we utilize a $512^3$ grid so as to reduce shot noise. For estimating the matter power spectrum, we use a $1024^3$ grid and the CIC kernel is deconvolved in Fourier space. 

\section{results}\label{sec:results}

Observe that the linear fluid equations~(\ref{eq:Euler})--(\ref{eq:Poisson}) can be combined into a single differential equation for the matter density, 
\begin{align} \label{ODE}
  \ddot \delta_{\rm cdm}+ {\cal H} \dot \delta_{\rm cdm} - 4 \pi G a^2   \bar \rho_{\rm cdm} \delta_{\rm cdm} =  4\pi G  a^2  \bar \rho_{\rm cdm} \deltaGR  \,. 
\end{align}
 We use the novel semi-analytical methods of \citet{Fidler:2016tir} 
to find the linear solution for $\delta_{\rm cdm}$ and its power spectrum,
 and confront it with the numerical solutions of the present approach.
We note that, for $\deltaGR =0$, equation~(\ref{ODE}) is the usual differential equation
for linear density fluctuations. The homogeneous solution of~(\ref{ODE}) is well known in the literature, so we will not report it here (see e.g.\ \citealt{Rampf:2015mza} and references therein).

\begin{figure} 
\includegraphics[width=\columnwidth]{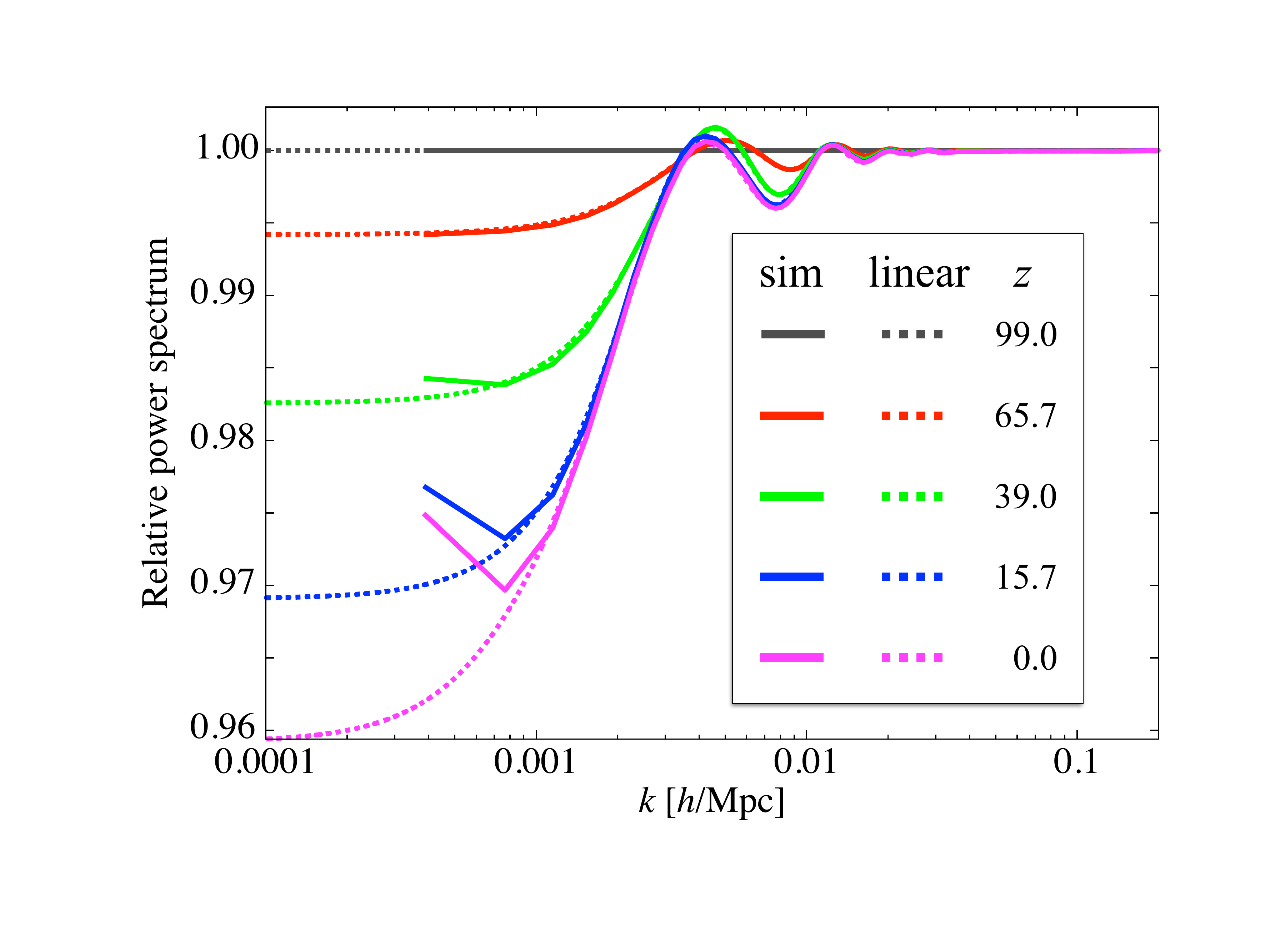}
\caption{Relative difference of the matter power spectrum $P(k)$  showing the impact of the relativistic perturbations $\deltaGR$ when switched on and off, i.e., $ P(\text{\small $\deltaGR = 0$}) / P(\text{\small $\deltaGR \neq 0$})$. All results are initialized at $z_{\rm ini} =99$. 
Shown are results from our numerical simulations (``sim''; solid lines) and linear theory predictions (``linear''; dotted lines) for the evolved differences of the power spectra at different redshifts. Linear predictions are obtained by applying the methods of \citet{Fidler:2016tir}.
} \label{fig:relativepower}
\end{figure}
In  Fig.\,\ref{fig:relativepower} we show the 
relative matter power spectrum when excluding/including GR perturbations, for various
redshifts. As evident, there is a striking agreement between the linear theory predictions (dotted lines) and our numerical results (solid lines).
On the scales we simulate, the impact of GR perturbations is negligible for $k > 10^{-2}\,h/$Mpc, but yields up to about $3\%$ corrections on larger scales. We note that
at very large scales, close to the fundamental mode of our simulations, our numerical results depart from the linear predictions; this is due to sample variance because of the lack of $N$-body particles, and thus could easily be rectified if needed.

\begin{figure}
\includegraphics[width=\columnwidth]{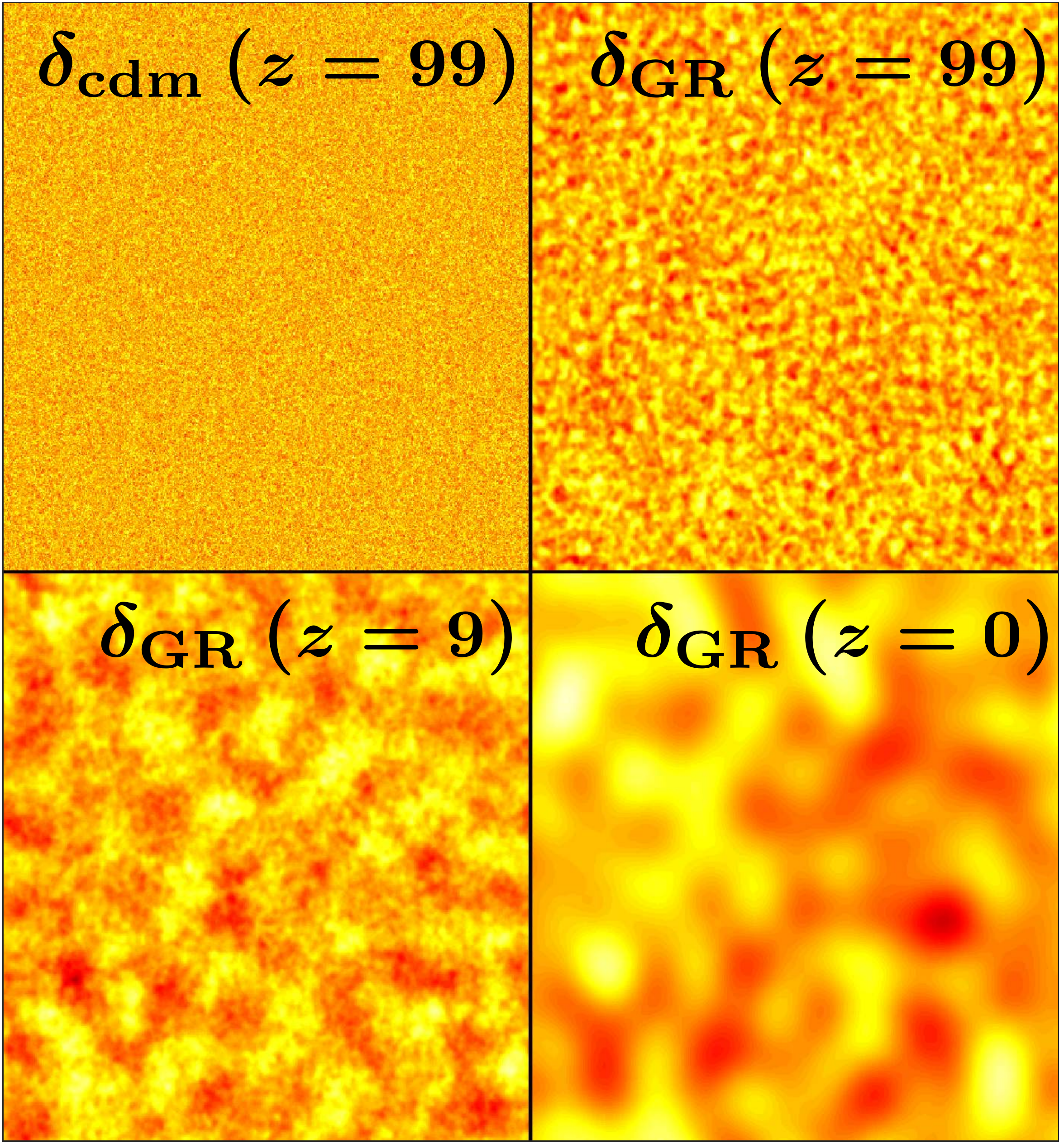}
\vskip-0.1cm\caption{Shown are slices of the effective density perturbations $\deltaGR$ (see equation~(\ref{def:deltaGR})) at various redshifts, compared against $\delta_{\rm cdm}$ at $z=99$. The CDM density does not change visibly at later times.
The thickness of the slices is 512\,Mpc$/h$ and the width is 16384\,Mpc$/h$. 
For better visibility we have increased the amplitudes of $\delta_{\text{\tiny GR}}(z)$ with respect to $\delta_{\rm cdm}( \text{\scriptsize $z=99$} )$ by factors of $1099$, $15265$, and $53936$ respectively for the redshifts $z=99,9$ and $0$. 
} \label{fig:density}
\end{figure}
In Fig.\,\ref{fig:density} we show slices of the GR perturbations $\delta_{\text{\tiny GR}}$ at various redshifts, compared to $\delta_{\rm cdm}$ at $z_{\rm ini}=99$.
The relative importance of the radiation and CDM perturbations scales as $\Omega_{\rm rad} / \Omega_{\rm cdm}$, and at the largest scales simulated at $z = 99$, the radiation transfer function is roughly $4/3$ times the CDM counterpart.
At smaller scales, the effect of {\it radiation free-streaming} 
is clearly visible, and becomes more pronounced at low redshift. Note that the largest structures in the box are not directly visible, since the density slices are dominated by scales around the peak of the power spectrum, $k \sim 0.01\,h/$Mpc. Since the radiation component has free-streamed out of these scales at $z=99$, the radiation and CDM slices look quite different at this redshift. 
The CDM density, by contrast, follows a nearly scale-invariant growth on the scales shown, and thus there is no visible evolution, which is the reason why we only show the slice at $z=99$. 

\begin{figure} 
\vskip0.01cm\includegraphics[width=\columnwidth]{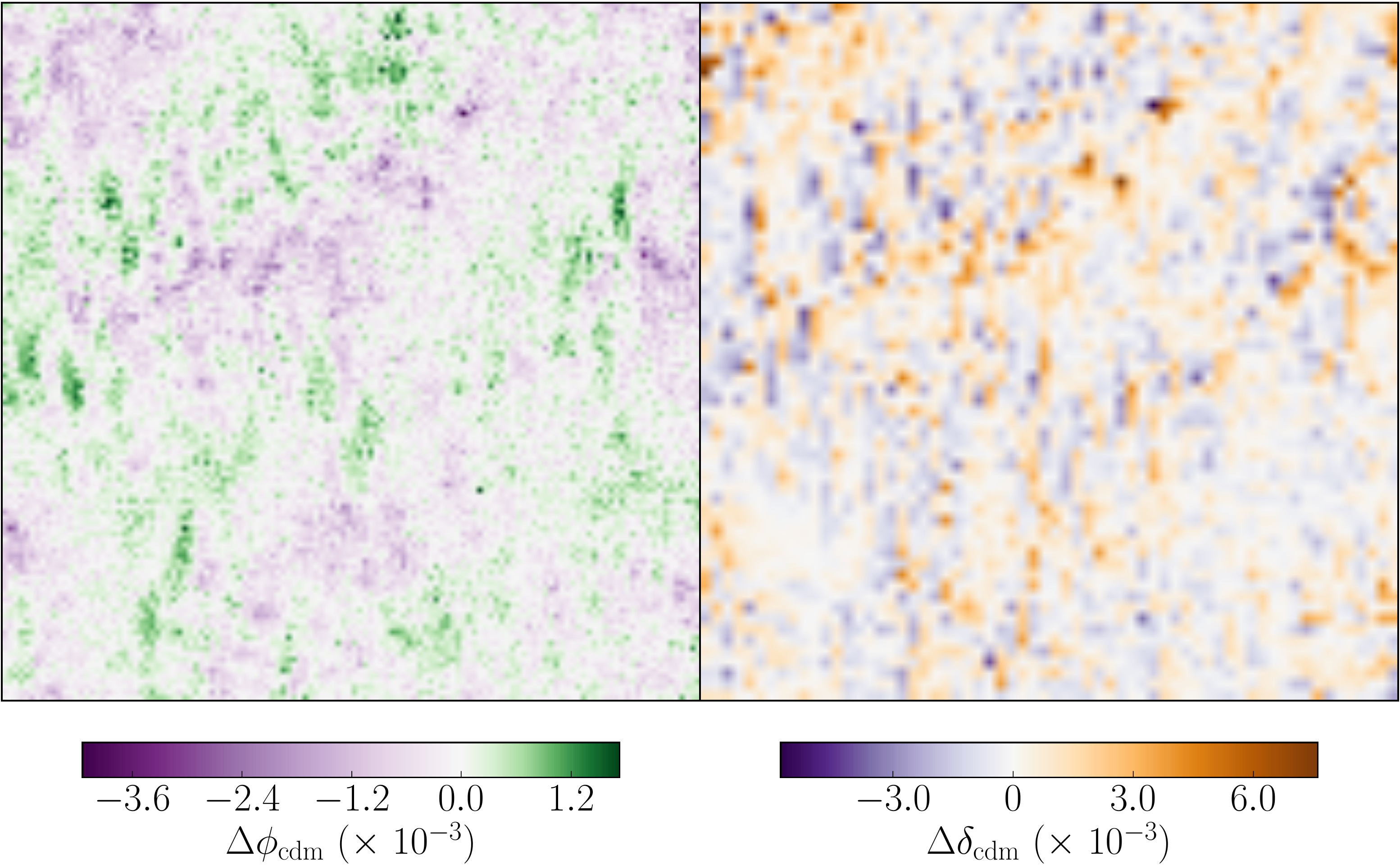}%
\caption{The left panel shows the difference  $\Delta \phi_{\rm cdm} \equiv \phi_{\rm cdm}(\text{\scriptsize $\deltaGR \neq 0$}) - \phi_{\rm cdm} (\text{\scriptsize $\deltaGR = 0$})$ of snapshots obtained from our hybrid simulations, where $\phi_{\rm cdm}$ is the potential of the scalar part of the Lagrangian displacement field ($\fett{\Psi} = \nabla \phi_{\rm cdm}$).
The right panel shows the difference of Eulerian density contrasts, $\Delta \delta_{\rm cdm} \equiv \delta_{\rm cdm}(\text{\scriptsize $\deltaGR \neq 0$}) - \delta_{\rm cdm} (\text{\scriptsize $\deltaGR = 0$})$. Both panels are evaluated at final time $z=0$, 
and show a region of $2048$\,Mpc/$h$ side length.
Clusters move (in Eulerian coordinates) wherever $\Delta\phi_{\rm cdm}$ exhibits high gradients. The two simulations predict the same structure in regions of constant $\Delta\phi_{\rm cdm}$.
} \label{fig:relativedifferencedelta}
\end{figure}

In Fig.\,\ref{fig:relativedifferencedelta} we show two panels, the left one depicts the difference of the Lagrangian potentials $\Delta \phi_{\rm cdm} \equiv \phi_{\rm cdm}(\text{\scriptsize $\deltaGR \neq 0$}) - \phi_{\rm cdm} (\text{\scriptsize $\deltaGR = 0$})$, and the right panel the difference of Eulerian density contrasts, $\Delta \delta_{\rm cdm} \equiv \delta_{\rm cdm}(\text{\scriptsize $\deltaGR \neq 0$}) - \delta_{\rm cdm} (\text{\scriptsize $\deltaGR = 0$})$. Both panels are evaluated at final time $z=0$.
From these panels, one can recognize that
large gradients in $\Delta \phi_{\rm cdm}$ cause clusters to move, and 
the change of final positions of these clusters manifest as  ``dipoles'' in the Eulerian panel.

In addition to the slight displacement of high-density regions in the presence of radiation/GR, we have investigated whether there are effects concerning the relative abundance of different cosmic web elements. One way to characterize such effects are the so-called mass filling fractions (MFF), which measure how much of the overall mass content is contained in clusters, filaments, sheets, and voids. (The MFF is defined such that the sum of the individual MFFs is unity.) Of course, the MFFs depend on the definition of such structures, and in the present work we have chosen the Lagrangian classifier \texttt{LICH} \citep{Leclercq:2016eur}. We find that radiation/GR has at most a ${\cal O}(10^{-5})$ relative effect on these numbers, leaving the MFFs effectively unchanged; in detail they are 
MFF({\small clusters}) $\simeq 0.097$, 
MFF({\small filaments}) $\simeq 0.414$,  
MFF({\small sheets}) $\simeq 0.389$, and 
MFF({\small voids}) $\simeq 0.100$. 
Thus, radiation/GR induces only a large-scale shift of cosmological structures (cf.\ Fig.\,\ref{fig:relativedifferencedelta}), but does not change their relative abundance.

\section{Summary and discussion}\label{sec:conclusion}

We have performed novel hybrid simulations 
to investigate cosmological structure formation for all fluid species within General Relativity. 
By using our developed code \texttt{COSIRA} (``COsmological SImulations with RAdiation''), we evolve photon and neutrino perturbations to first-order within cosmological perturbation theory, whilst solving for the matter perturbations in full non-linearity in Newtonian gravity.

Relativistic computations within \texttt{COSIRA} are performed in the $N$-body gauge \citep{Fidler:2015npa}, which is a gauge constructed such that at sufficiently late times the equations of motion for CDM coincide exactly with the ones in Newtonian gravity. In that gauge also the relativistic spatial volume element coincides at all times with the one in Newtonian gravity, which makes interpreting mass densities and velocites in the relativistic simulation a trivial exercise.
We also note that at sufficiently late times we have for the line element in the $N$-body gauge, equation~(\ref{lineNbody}), that $A = 0$ and $H_{\rm T} = 3 \zeta$, where $\zeta$ is the gauge-invariant comoving curvature perturbation (see e.g.\ \citealt{Fidler:2016tir}). This should enable straightforward post-analyses of the output of our hybrid simulations on a fully relativistic space-time,  e.g., to investigate weak lensing via ray tracing techniques on the $N$-body output.

Our numerical code \texttt{COSIRA} delivers extremely stable results, especially on large scales. This is most clearly seen in Fig.\,\ref{fig:relativepower} where we compare our numerical findings against theoretical predictions. The agreement between $N$-body simulation and linear theory results is excellent, except on very large scales, where our numerics are mostly affected by particle shot noise. It is because of this agreement that the non-linear mode coupling induced through the first-order GR corrections seems to be highly suppressed, however further studies with higher precision are required to assess this in more detail.

In the present case study we have focused on the regime of cosmological structure formation starting from $z_{\rm ini}=99$ until today. \texttt{COSIRA} makes it however possible to initialize relativistic simulations also at earlier times, possibly even as early as recombination. 
Furthermore, our findings enable us to perform comparisons to other relativistic simulations, where different approximations and assumptions have been applied, see e.g.\ \citet{Adamek:2015eda,Bentivegna:2015flc,Hahn:2016roq}.
The limitation of the present approach is essentially by two effects, which are of second order in cosmological perturbation theory. First, the radiation source is computed at the linear order using \texttt{CLASS}. Higher-order corrections are expected to be very small, especially since the source is only relevant at early times. Second, our approach is based on the $N$-body gauge, 
but so far this gauge is only defined at the linear order. Relativistic corrections to the CDM evolution are expected at second order in the metric potentials, and are typically negligible. These corrections however could matter when studying observables that are suppressed at the linear order, such as the relativistic matter bispectrum \citep{Tram:2016cpy}.

Finally, we note that in the present paper, for reasons of better comparison with the theoretical model, we have chosen to initialize our simulations with the actual matter power spectrum at $z_{\rm ini}$, for both runs when including and excluding radiation. While this procedure is perfectly justified in the case when radiation is included, it is rather not when neglecting radiation perturbations. Indeed, in the latter case, it is a very common practice to initialize $N$-body simulations by using today's matter power spectrum, and rescale its amplitude by the growing mode of linear density fluctuations back to when the simulation is initialized (e.g., \citealt{Schneider:2015yka,Zennaro:2016nqo}). 
This method, sometimes referred to as ``back scaling'',  
produces a matter power spectrum at $z_{\rm ini}$ of a designed universe with vanishing radiation content.
Of course, the back scaling method neglects by construction the evolution of radiation (see Fig.\,\ref{fig:density}). This and many more details should be assessed in forthcoming studies and confronted with the present approach to sufficient precision.

\section*{Statement of contribution}

JB implemented the general relativistic modification into the $N$-body code \texttt{GADGET-2}. CR was involved in the conception and design of this project, and wrote the paper. TT modified the Einstein--Boltzmann code \texttt{CLASS} such that it outputs the relativistic correction function $\gamma$ and other related quantities. 
FL produced the Eulerian and Lagrangian plots (Fig.\,\ref{fig:relativedifferencedelta}) and performed the cosmic web analysis using \texttt{LICH}.
CF and SH contributed to the interpretation of the results, and to the overall clarity of this paper. All authors read and approved the final manuscript.

\newpage

\section*{Acknowledgements}

We thank Julian Adamek for useful discussions.
JB and SH acknowledge support from the Villum Foundation.
CR~acknowledges the support of the individual fellowship RA 2523/1-2 from the Deutsche Forschungsgemeinschaft (DFG).
FL acknowledges funding from the European Research Council through grant 614030, Darksurvey.
CF is supported by the Wallonia-Brussels Federation grant ARC11/15-040 and the Belgian Federal Office for Science, Technical \& Cultural Affairs through the Interuniversity Attraction Pole P7/37. 



\end{document}